\documentclass[aip,sd,amsmath,amssymb,reprint]{revtex4-1}

\usepackage{graphicx}
\usepackage{dcolumn}
\usepackage{bm}
\usepackage{epstopdf}
\usepackage{xcolor}
\usepackage{natbib}
\usepackage{subfigure} 
\begin{document}

\title{Enhanced proton acceleration in an applied longitudinal magnetic field}


\author{A. Arefiev}
\email{alexey@austin.utexas.edu}
\affiliation{Center for High Energy Density Science, The University of Texas, Austin, Texas 78712, USA}

\author{T. Toncian}
\affiliation{Center for High Energy Density Science, The University of Texas, Austin, Texas 78712, USA}

\author{G. Fiksel}
\affiliation{University of Michigan, 2200 Bonisteel Boulevard, Ann Arbor, Michigan 48109, USA}

\date{\today}

\begin{abstract}
Using two-dimensional particle-in-cell simulations, we examine how an externally applied strong magnetic impacts proton acceleration in laser-irradiated solid-density targets. We find that a kT-level external magnetic field can sufficiently inhibit transverse transport of hot electrons in a flat laser-irradiated target. While the electron heating by the laser remains mostly unaffected, the reduced electron transport during proton acceleration leads to an enhancement of maximum proton energies and the overall number of energetic protons. The resulting proton beam is much better collimated compared to a beam generated without applying a kT-level magnetic field. A factor of three enhancement of the laser energy conversion efficiency into multi-MeV protons is another effect of the magnetic field. The required kT magnetic fields are becoming feasible due to a significant progress that has been made in generating magnetic fields with laser-driven coils using ns-long laser pulses. The predicted improved characteristics of laser-driven proton beams would be critical for a number of applications. 
\end{abstract}

\maketitle


\section{Introduction} \label{Sec-Intro}

After first capturing the attention of the scientific community over a decade ago, laser-driven ion acceleration and now also its applications continue to remain an active area of experimental and theoretical research. One of the key appeals of the laser-driven ion acceleration is the compactness of the acceleration region. The accelerating fields sustained during laser-matter interactions at relativistic intensities are orders of magnitude higher than those used in conventional accelerators, reaching up to TV/m in strengths. These fields are highly transient, with a typical lifetime in the picosecond range, so the resulting energies of accelerated ions are typically in the range of several tens of MeV. 

Laser-driven proton beams are already being successfully used to produce high energy density matter~\cite{Patel2003} and to radiograph transient processes~\cite{Borghesi2002}. They also hold a great promise for such diverse applications as fast ignition in the energy production scheme by inertial confinement fusion~\cite{Roth2001}, tumor therapy \cite{Bulanov2002}, production of short-lived positron emitters for positron emission tomography \cite{Spencer2001}, and the brightness increase in conventional accelerators. However, in order to fully realize the potential of laser-driven proton and ion sources, a number of issues still remain to be addressed. The most pressing issues, depending on the application, are: 1) further increasing the proton energies, 2) controlling the proton beam divergence, 3) reducing the energy bandwidth of accelerated protons, and 4) increasing the conversion efficiency of the laser energy into the energy of the ion beam. Finding constructive solutions even to some of these issues would thus be extremely valuable. 

 \begin{figure}[hb]
	\centering
	\includegraphics[width=0.9\columnwidth]{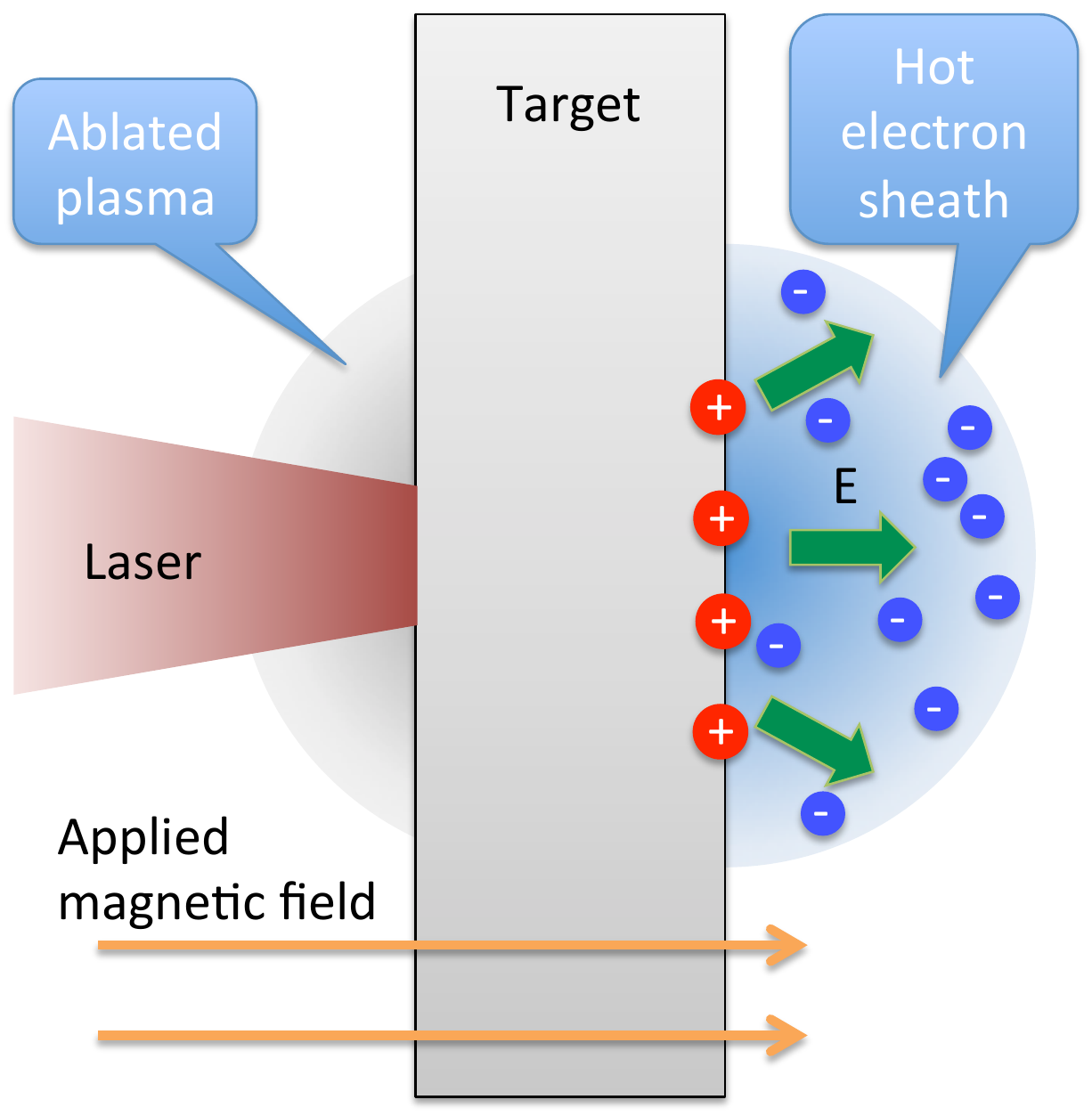} 
  \caption{Key features of the target normal sheath acceleration regime. Arrows at the bottom show the direction of an external magnetic field that can be applied to alter the ion acceleration both directly and by affecting the electron dynamics.} \label{Figure_TNSA}
\end{figure}

Recently, a significant progress has been made in generating strong magnetic fields with laser-driven coils using ns-long laser pulses~\cite{Gao2016,Law2016}. In the context of proton and ion acceleration, such fields can be viewed as static and nearly uniform. Such an external magnetic field applied along the laser propagation and normal to the irradiated target in proton acceleration experiments (see Fig.~\ref{Figure_TNSA}) offers a previously unexplored option of slowing down transverse electron transport in the target. Controlling the electron spread would control the proton acceleration, since the protons are coupled to the laser pulse only through the field generated by the electrons. In this paper, we examine this specific scenario using two-dimensional (2D) particle-in-cell (PIC) simulations with the goal to determine how an externally applied magnetic field impacts proton energy gain and the divergence of the proton beam. 

We find that a kT-level external magnetic field can sufficiently inhibit transverse transport of hot electrons in a flat laser-irradiated target. While the electron heating by the laser remains mostly unaffected, the reduced electron transport during proton acceleration leads to an enhancement of the maximum proton energies and the overall number of energetic protons. The resulting proton beam is extremely well collimated compared to a beam generated without applying a kT-level magnetic field. A factor of three enhancement of the laser energy conversion efficiency into multi-MeV protons is another important effect of the external magnetic field.

\begin{figure*}
	\centering
	\includegraphics[width=1.9\columnwidth]{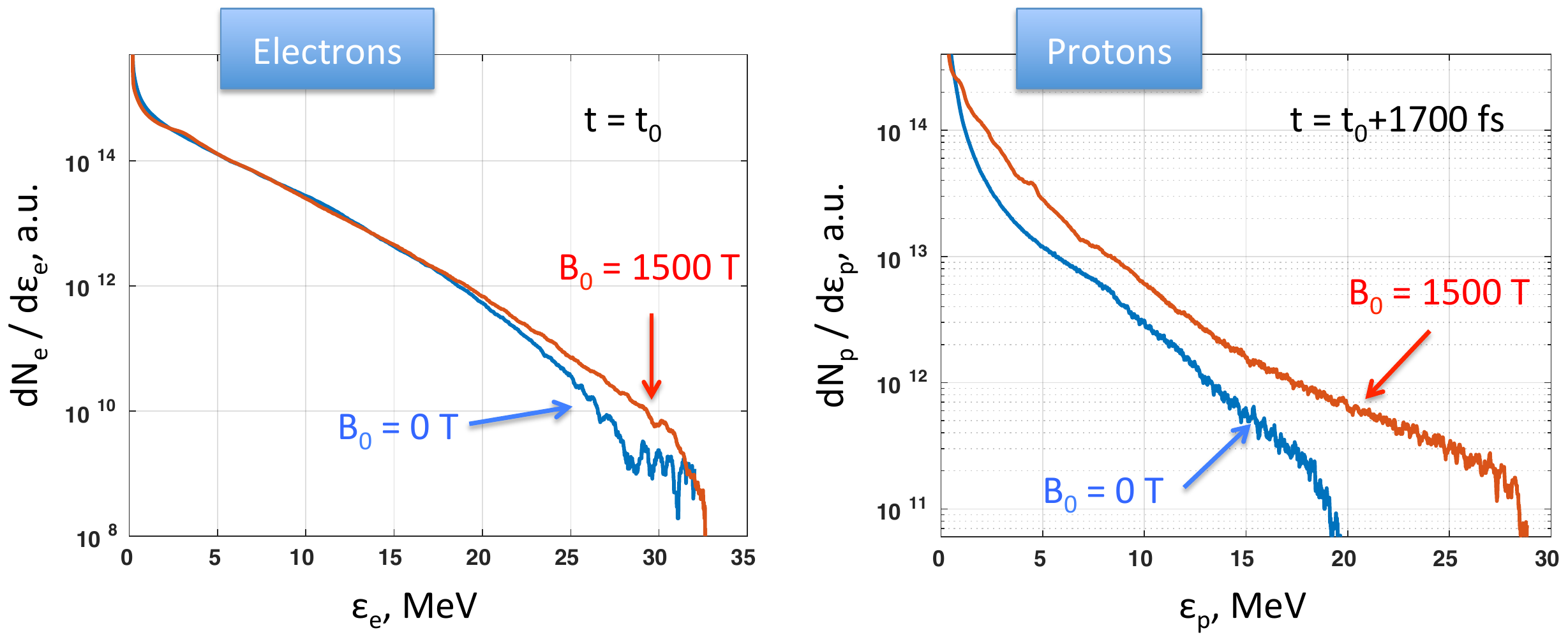} 
	  \caption{Snapshots on electron (left) and proton (right) spectra with and without the magnetic field. The electron snapshots are taken while the laser is interacting with the main target, whereas the proton snapshots are taken following an extended target expansion.} \label{Figure_edf}
\end{figure*}

\section{Background} \label{Sec-B}


One of the widely-used approaches to ion and proton acceleration is the so-called Target Normal Sheath Acceleration (TNSA) regime~\cite{Wilks2001}. In what follows, we provide a brief overview of its key features and summarize some of the key relevant experimental and theoretical results.

In the TNSA regime, a flat solid-density target is irradiated by an intense laser pulse (see Fig.~\ref{Figure_TNSA}) to generate hot electrons whose collisional mean-free-path greatly exceeds the thickness of the target. These electrons set up a sheath electric field at the surface of the target that prevents them from escaping. As a result of their large mean-free-path, the energetic electrons are effectively decoupled from the bulk of the electrons, which remains cold, and are able to retain their energy while traversing the target. The sheath field that keeps the hot electrons in the target gradually accelerates protons or ions off the surface of the target, hence the TNSA name, eventually producing an energetic proton or ion beam. 

Assuming a sharp plasma-vacuum interface, the sheath electric field, that energetic electrons with density $n_H$ and characteristic kinetic energy $T_H$ set up at the rear side, can be estimated as 
\begin{equation} \label{sheath_field_0}
	E \sim \sqrt{n_H T_H}.
\end{equation}
The characteristic scale of the sheath is the Debye length $\lambda_D = \sqrt{T_H / 4 \pi n_H e^2}$, where $e$ is the electron charge. The estimate given by Eq.~(\ref{sheath_field_0}) implies that that the sheath electric field traps electrons with the characteristic energy $T_H$ inside the target, i.e., $|e| E \lambda_D \sim T_H$. It is evident from Eq.~(\ref{sheath_field_0}) that whether a strong sheath field can be generated or not is determined by the ability of the laser pulse to generate copious energetic electrons. It has been well recognized that electron acceleration in the laser-plasma interaction at the front surface of the target is the key to effective proton acceleration and this is one of the motivating factors for the ongoing active electron acceleration research~\cite{Peebles2016,Sorokovikova2016,Arefiev2016}. The strength of the sheath electric field is given by Eq.~(\ref{sheath_field_0}) only if the characteristic scale of the plasma density $l$ is less than the Debye length $\lambda_D = \sqrt{T_H / 4 \pi n_H e^2}$, where $e$ is the electron charge. If $l \gg \lambda_D$, then the strength of the sheath electric field becomes greatly reduced compared to that given by Eq.~(\ref{sheath_field_0}). That is why a sharp plasma-vacuum boundary is a must for efficient proton acceleration.

Once the sheath electric field is set up at the plasma boundary at the rear side of the target, it initiates a gradual plasma expansion. The key physics elements of this regime are essentially the same as for the plasma expansion into a vacuum, which was first theoretically considered in an early work by Gurevich et al. \cite{Gurevich}. The expansion enables an energy transfer from the hot electrons to the ions via an evolving electric field in the expanding plasma flow. 

P. Mora has extended the model proposed by Gurevich et al. to obtain accurate results concerning the structure of the ion front, the resultant ion energy spectrum, and, most importantly, the maximum ion energy~\cite{Mora2003}. The extended model predicts an exponential shape of the ion energy spectrum with a time-dependent cutoff. The cut-off energy $\varepsilon_{\max}$ increases with the acceleration time $t$ as 
\begin{eqnarray}
	\varepsilon_{\max} \propto \left( \ln[ (2/\mbox{e})^{1/2} \omega_{pi} t ] \right)^2,
\end{eqnarray}
where e is the Euler's number, $\omega_{pi} = \sqrt{4 \pi n_i q_i^2 / m_i}$ is the ion plasma frequency, and $q_i$ and $m_i$ are the ion charge and mass. In this model, the characteristic time scale, $1/\omega_{pi}$, is set by the ion dynamics. The model was later further improved by P. Mora by including adiabatic cooling of electrons~\cite{Mora2005}. The salient result of this later work is a pronounced double-layer structure of the ion front.

The double-layer electric field structure responsible for the proton acceleration was first experimentally discovered by Romagnani et al. \cite{Romagnani2005} in a study investigating generation of multi-MeV protons at the rear surface of a thin solid foil irradiated by an intense  short-pulse laser. The structure of the electric field driving the plasma expansion has been imaged with high spatial and temporal resolution using transverse proton probing. The main features of the experimental observations, namely an initial strong sheath field and the late-time electric field peaking at the front of the proton beam, are consistent with the theoretical predications for the plasma expansion into vacuum.

In 2008, J. Psikal et al. demonstrated using PIC simulations~\cite{Psikal2008} that the maximum energy of fast protons depends on the cross section of the hot electron cloud behind the target. Therefore, the use of small targets with all dimensions less or comparable to the laser spot size can be beneficial. Such targets enhance the efficiency of laser energy transformation into fast protons by reducing the spread of hot electrons in the transverse plane. J. Psikal et al. have considered targets that have different shapes and found that a cylindrical target enhances the maximum proton energy. However, this target also produces an undesirable divergence of the proton beam, which leads to lower densities of fast protons.

The first experimental demonstrations of the enhanced proton acceleration in solid foils of limited transverse extent using 100 TW class lasers were reported by S. Buffechoux et al.~\cite{Buffechoux2010} and T. Toncian et al.~\cite{Toncian2011}. The reduced target surface area in these experiments allowed for transverse electron refluxing, with the hot electrons being reflected from the target edges during or shortly after the laser pulse. The refluxing maintains a hotter, denser, and more homogeneous electron sheath around the target for a longer time. Compared to the targets without the refluxing, the measured maximum proton energies and laser-to-ion conversion efficiencies were considerably increased. 

More recently, Margarone et al.~\cite{Margarone2012} showed that the proton acceleration and the resulting energy can be further increased by employing targets that enhance the absorption of the incident laser beam. In their experiments, Margarone et al.~\cite{Margarone2012} used targets with a monolayer of polystyrene nanospheres on the front side whose diameter was less than the laser wavelength. Using a 100 TW 805 nm laser, the maximal proton energy was increased by about 60\% compared to a planar foil for an optimal spheres' diameter of 535 nm.

The described successful research on TNSA proton acceleration has stimulated a wide range of concepts that aim to utilize TNSA proton beams to solve some of the outstanding technological challenges. One of such concepts that is worth emphasizing is the proton-driven fast ignition for inertial confinement fusion~\cite{Roth2001}. As suggested in 2001 by M. Roth et al.~\cite{Roth2001}, an intense TNSA proton beam can be used to ignite a pre-compressed cold fuel sphere. This approach would significantly increase the gain, reduce the driver energy, and relax the symmetry requirements for compression. 

In the scenario proposed by M. Roth et al., multiple petawatt-class laser beams are focused onto a target which is spherical in shape. Since the TNSA mechanism accelerates protons predominantly normal to the target surface, using a target with a curved rear (non-irradiated) surface makes it possible to ballistically focus protons to a small spot. The feasibility of focusing a proton beam by employing a laser-irradiated spherical shell was demonstrated for the first time in 2003 by P. Patel et al.~\cite{Patel2003}.  Details of the proton focusing process have been studied by S. Kar et al. \cite{Kar2011} and D. Offermann et al. \cite{Offermann2011}. Both studies concluded that the proton focusing is intrinsically chromatic. The underlying cause is the shape of the expanding TNSA sheath field front~\cite{Romagnani2005} that leads to a varying divergence as a function of the proton energy in the beam.

 \begin{figure*}[htb]
	\centering
	\subfigure{\includegraphics[width=0.9\columnwidth]{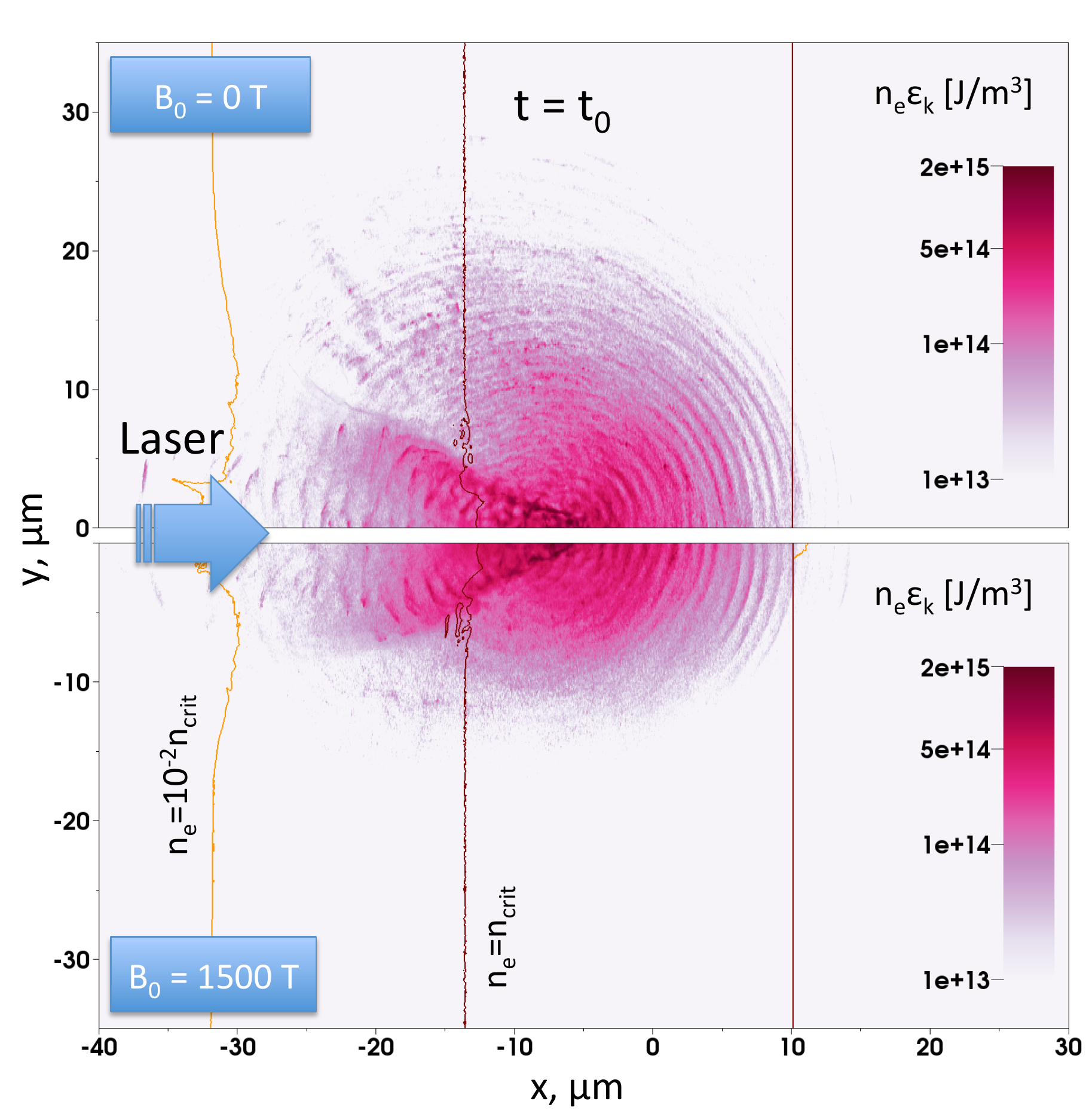} }
	\subfigure{\includegraphics[width=0.9\columnwidth]{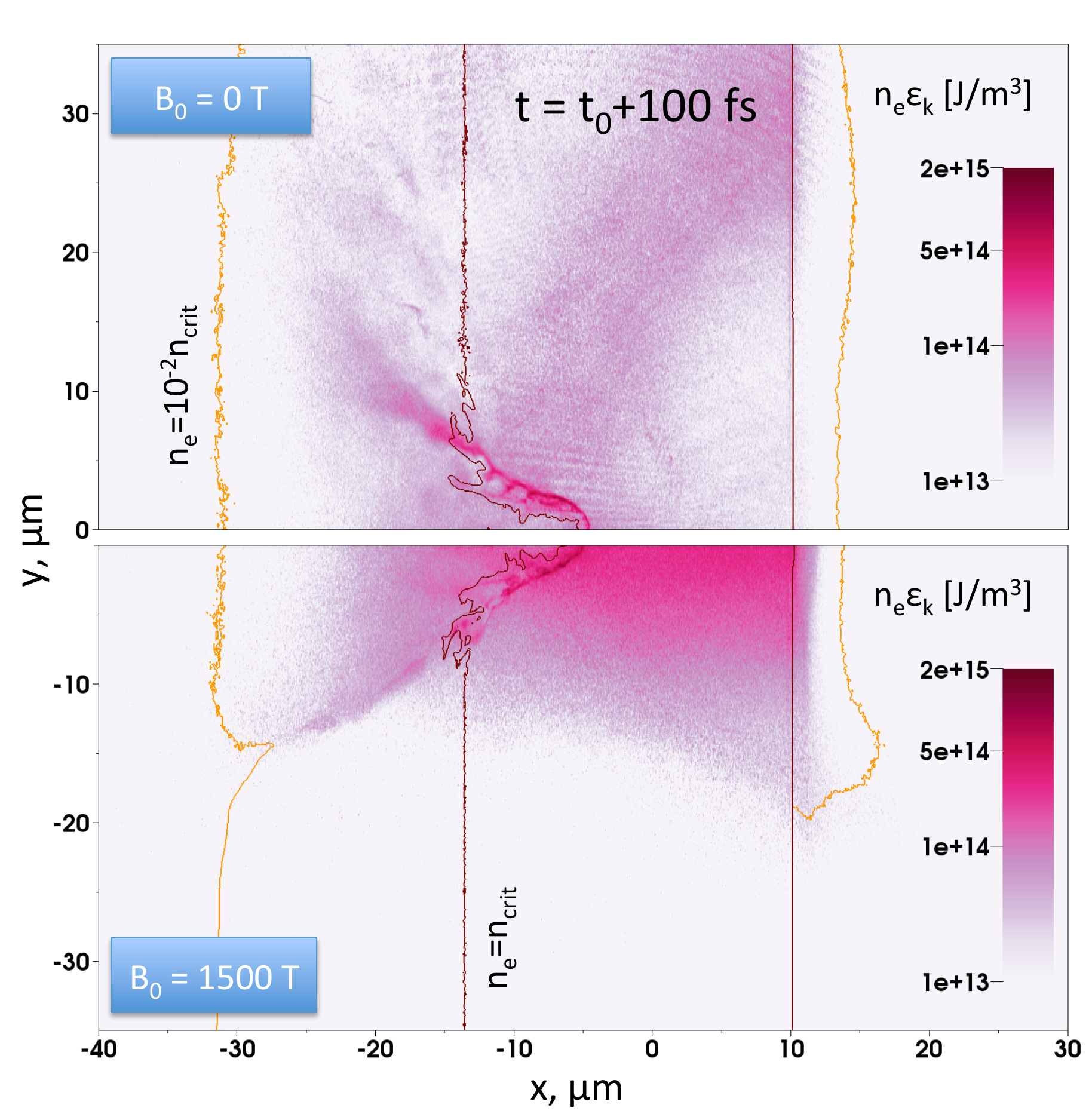}}
  \caption{Snapshots of electron energy density with (lower two panels) and without (upper two panels) an applied longitudinal magnetic field. The contours show electron density surfaces in the target corresponding to $n_e = n_{crit}$ and $n_e = 10^{-2} n_{crit}$. At $t = t_0$ the laser pulse is interacting with the main target.} \label{Figure_ek}
\end{figure*}

As pointed out in the Introduction (Sec.~\ref{Sec-Intro}), an externally applied magnetic field that can be generated using laser-driven coils~\cite{Daido,Daido87,Fujioka} offers a previously unexplored ``control knob'' for TNSA proton beams. K. Mima et al. have recently shown that an external kT-level magnetic field can be beneficial for electron fast ignition by improving the coupling efficiency into the fusion core~\cite{Mima2016}. In the following Sections we analyze by means of two-dimensional numerical simulations the effect of such a magnetic field on the TNSA of protons at moderate laser intensities.

\section{Simulation setup} \label{Sec-Setup}

We examine the role of an applied magnetic field using fully self-consistent two-dimensional (2D) particle-in-cell simulations performed with an open-source code EPOCH~\cite{Arber2015}.

We simulate only the interaction with the main part of the laser pulse. However, a pre-pulse is likely to be present in a typical experiment~\cite{Schollmeier2015}. Such pre-pulse can deliver enough energy to ionize a solid density target, causing it to pre-expand. In our simulations, we account for the pre-pulse by initializing the target as a plasma with a given pre-plasma at the surface irradiated by the laser pulse. Specifically, an initial electron density profile is given by
\begin{equation}
n_e =
   \begin{cases}
      n_0 \exp[-x^2/ L^2] ,  & \text{for $x <  x_L$;}\\
      n_0,  & \text{for $x_L \leq x \leq  x_R$;}\\
      0,  & \text{for $x >  x_R$,}
   \end{cases} \label{n_sim}
\end{equation}
where $x_L = 0$ $\mu$m, $x_R = 10$ $\mu$m, and $L = 4$ $\mu$m. We set $n_0 = 50 n_{crit}$, where $n_{crit}$ is the critical density,
\begin{equation} \label{nc}
n_{crit} \equiv   \frac{m_e \omega^2}{4 \pi e^2},
\end{equation}
for which the electron plasma frequency $\omega_{pe} = \sqrt{4 \pi n_e e^2 / m_e}$ becomes equal to the frequency of the laser pulse, where $m_e$ and $e$ are the electron mass and charge. For simplicity, we assume that the target ions are protons with the same initial density profile as the electrons. The electron population is initialized using 150 macro-particles per cell. In order to improve the statistics for the protons accelerated at the rear side of the target, we initialized the proton distribution using 25 macro-particles per cell for $x \leq 8.5$ $\mu$m and using 500 macro-particles per cell for $x > 8.5$ $\mu$m. All electrons and protons are initially cold. The size of the domain in the simulation is 200 $\mu$m (6000 cells) along the $x$-axis and 300 $\mu$m (3000 cells) along the $y$-axis. 

We use a laser pulse propagating along the $x$-axis, whose focal plane in the absence of plasma is located at $x = 0$ $\mu$m. The laser fields have a Gaussian profile along the $y$-axis, with the electric and magnetic fields in the focal plane given by
\begin{equation}
\left. E_{wave} \right/ E_0 = \left. B_{wave} \right/ B_0 = S(t) \exp \left( - \left. y^2 \right/ w_0^2 \right),  
\end{equation}
where $w_0 = 6.76$ $\mu$m and the temporal profile is 
\begin{equation}
S(t) =
   \begin{cases}
      \sin^2(\pi t/ T) ,  & \text{for $t <  T$;}\\
      0,  & \text{for $t \geq  T$},
   \end{cases} \label{E_sim}
\end{equation}
with $T = 150$ fs. The laser wavelength is $\lambda = 1$ $\mu$m and the peak laser intensity at $x = y = 0$ $\mu$m is $I = 2.2 \times 10^{19}$ W/cm$^2$, which corresponds to $a_0 = 4$. The normalizing amplitudes $E_0$ and $B_0$ are maximum electric and magnetic fields in a plane wave with $a_0 = 4$. In all our simulations, the laser electric field has $x$ and $y$ components, whereas the laser magnetic field has only a $z$-component. 

We also initialize a uniform magnetic field $B_0$ directed along the $x$-axis. All the fields evolve according to Maxwell's equations based on the initial conditions. This field represents a quasi-static externally applied magnetic field in experiments. 



\begin{figure*}
	\centering
	\includegraphics[width=1.9\columnwidth]{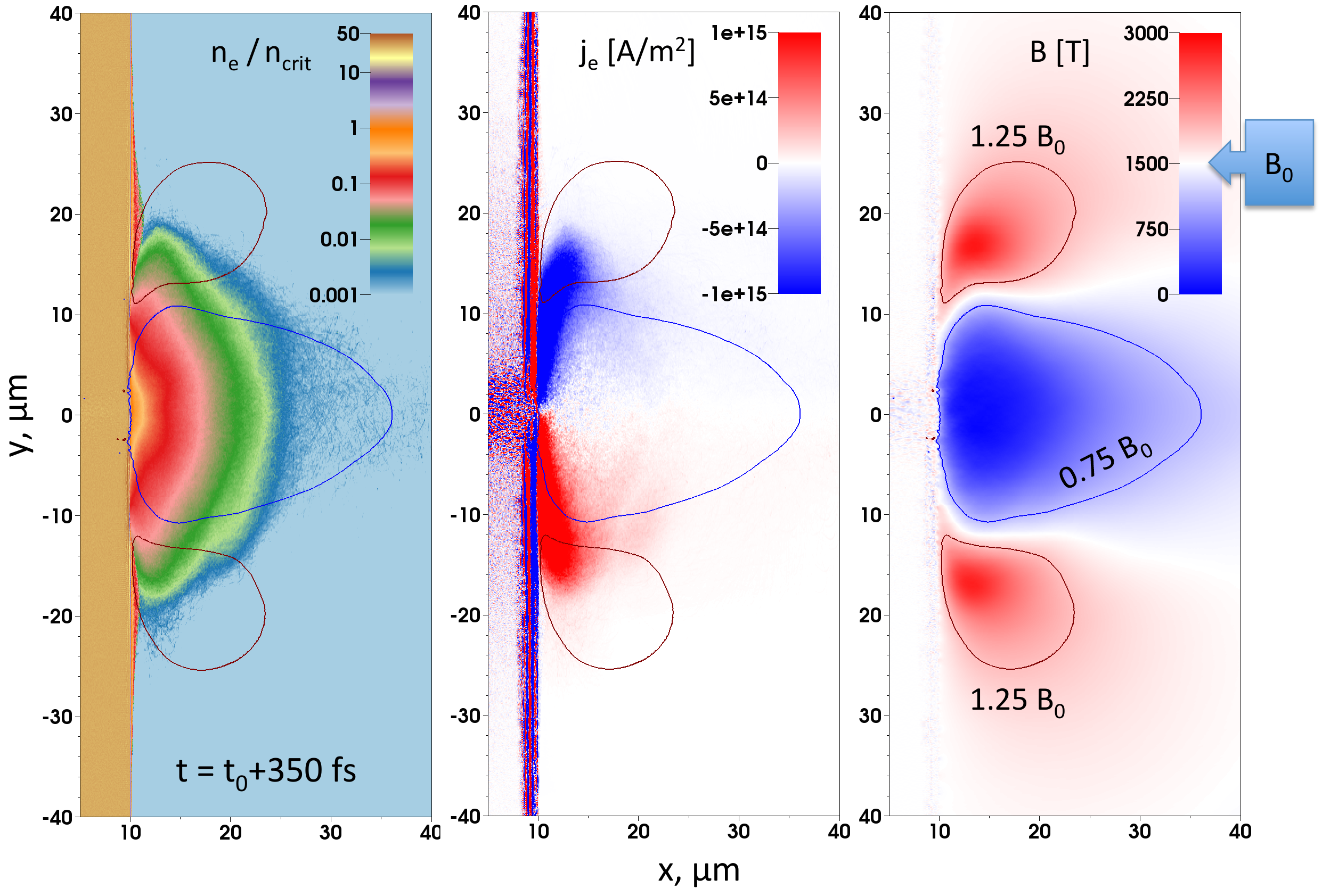} 
	  \caption{Snapshots of electron density (left), electron current $j_z$ (center), and longitudinal magnetic field $B_x$ (right) at the rear-side of the target 350 fs after the laser-target interaction. The red and blue contours in all three panels correspond to $B_x = 1.25 B_0$  and $B_x = 0.75 B_0$, where $B_0 = 1.5$ kT is the strength of the applied magnetic field. } \label{Figure_sheath}
\end{figure*}

\section{Simulation results} \label{Sec-Setup}

In order to examine the impact of an externally applied magnetic field on proton acceleration, we have performed simulations for three different values of the magnetic field, $B_0 = 0$ T, 150 T, and 1.5 kT. All other simulation parameters were the same in these three runs. Our results show that the 150 T magnetic field is not sufficient to impact the proton spectrum either by affecting the electron heating during the laser-target interaction or by affecting how the heated electrons accelerate the protons. However, we find that the 1.5 kT magnetic field has a profound impact on the resulting proton beam as compared to the regime without an applied magnetic field ($B = 0$ T). In what follows, we therefore focus on comparing the physics between the case with $B_0 = 1.5$ kT and the case with $B_ 0 = 0$ T. From now on, we refer to the them as the cases with and without the magnetic field.

\begin{figure*}[htb]
	\centering
	\includegraphics[width=2.1\columnwidth]{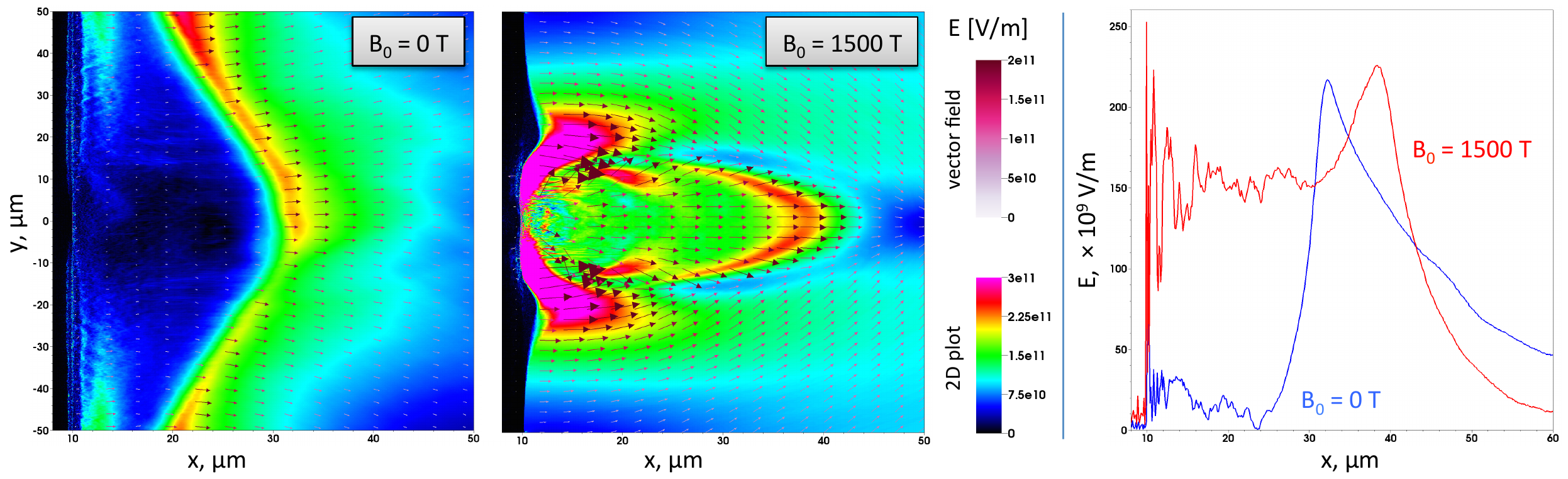} 
	  \caption{Snapshots of the electric field at $t = t_0 + 0.6$ ps for the case with and without the applied magnetic field. The right panel shows the line-outs along the $x$-axis ($y$ = 0 $\mu$m) for the plots shown in the left and central panels.} \label{Figure_field}
\end{figure*}

The left panel in Fig.~\ref{Figure_edf} shows a comparison of electron energy spectra during the laser-target interaction with and without the magnetic field. The snapshots were taken 500 fs into the simulation, so, for convenience, we define $t_0 = 500$ fs. The spectra look virtually indistinguishable below 20 MeV. There is only a slight enhancement in the number of most energetic electrons in the presence of the magnetic field. This is not very surprising, because the electron cyclotron frequency in this case is much lower than the frequency of the laser. The cyclotron frequency, $\omega_{ce} \equiv |e| B_0 / m_e c$, in a $B_0 = 1.5$ kT field is roughly seven times lower than the frequency of a laser $\omega$ with wavelength $\lambda = 1$ $\mu$m, where $c$ is the speed of light. In principle, it is possible for a longitudinal magnetic field to dramatically enhance the energy gain of electrons accelerated by an intense laser pulse even if $\omega_{ce} \ll \omega$, but this requires additional longitudinal electron pre-acceleration and an extended acceleration length~\cite{Arefiev2015}. Both of these conditions have obviously not been met in our case. 


The right panel in Fig.~\ref{Figure_edf} shows a comparison of proton energy spectra 1.7 ps after the laser-target interaction. These are the protons accelerated by electric fields generated by the hot electron populations shown in the left panel of Fig.~\ref{Figure_edf}. Based on the the slight enhancement of the electron energy spectrum, one would not expect significant changes in the proton spectrum. However, the right panel in Fig.~\ref{Figure_edf} tells a different story. In the presence of the 1.5 kT magnetic field, there is a considerable enhancement in the total number of accelerated protons over a wide range of energies. The total number of protons with energies above 1 MeV increased by a factor of 2.25, while the number of protons with energies above 10 MeV increased by a factor of 2.83. This translates into a considerable enhancement in efficiency of the energy conversion from the laser pulse into energetic protons. The energy content in the proton distribution with energies above 1 MeV and 10 MeV increased by factors 2.4 and 3.25, respectively. The cutoff proton energy also increases significantly, roughly by 50\%. 

This result indicates that a longitudinal magnetic field can enhance proton energies without enhancing the energetic part of the electron spectrum responsible for generating an accelerating sheath field. This is a welcome development, because a magnetic field much stronger than 1.5 kT would be needed to satisfy the condition  $\omega_{ce} \sim \omega$ and thus dramatically enhance the acceleration in a laser pulse for a large number of energetic electrons.

In order to determine the reason for the enhancement of the proton spectrum, we take a closer look at the dynamics of energetic electrons produced as a result of a laser-target interaction at the front surface of the laser-irradiated target. Figure~\ref{Figure_ek} shows snapshots of the electron energy density with and without the magnetic field. The snapshots in the two left panels were taken at the same instant as the snapshots of the electron distribution in Fig.~\ref{Figure_edf}. The energetic electrons generated on the left side of the target (the laser-irradiated side) are injected into the target and move to the right beyond the critical surface, with the magnetic field reducing the divergence of the injected electron beam. The right two panels in Fig.~\ref{Figure_ek} show the electron energy density 100 fs later. A relativistic electron can travel approximately 30 $\mu$m during this time period. As a result, the energetic electrons in the absence of the magnetic field spread in the lateral direction. In contrast to that, the 1.5 kT magnetic field stems the lateral electron spread. This effect can also be seen from the $n_e = 10^{-2} n_{crit}$ contour at the rear side of the target. This contour effectively outlines the electron sheath. The sheath is laterally localized in the presence of the magnetic field, whereas there is no visible localization without the magnetic field. 

The charge separation in the electron sheath at the rear side of the target generates a longitudinal electric field directed away from the target surface. This field prevents energetic electrons from leaving the target and, at the same time, it initiates a target expansion. Figure~\ref{Figure_sheath} shows the electron density profile in the presence of the magnetic field another 250 fs later. The light-green color marks electron densities in the range of $10^{-2} n_{crit}$. By comparing this contour to the $n_e = 10^{-2} n_{crit}$ contour in the lower-right panel of Figure~\ref{Figure_ek}, we find that the plasma plume is primarily expanding in the longitudinal direction, while the lateral expansion on this time scale is essentially negligible.

\begin{figure*}[htb]
	\centering
	\includegraphics[width=1.6\columnwidth]{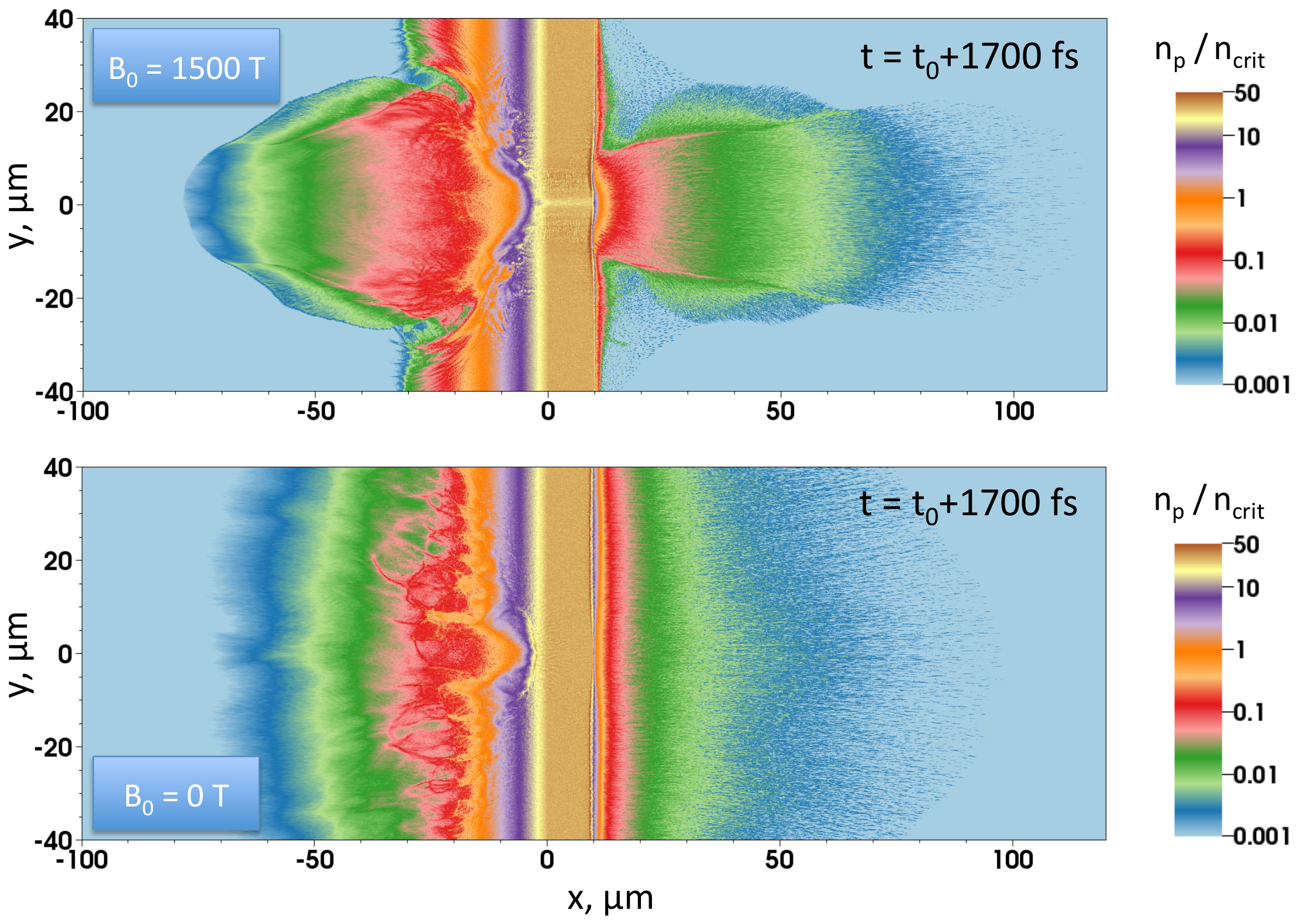} 
	  \caption{Snapshots of proton density profiles averaged over ten laser periods for $B_0 = 1.5$ kT (upper panel) and $B_0 = 0$ T (lower panel). The snapshots were taken at 1.4 ps after the start of the simulation. Proton densities are normalized to the critical density $n_c$ and color-coded on a log-scale.} \label{Figure_ions}
\end{figure*}

There is however a transverse electric field that is directed outwards on both sides of the expanding plasma plume. This electric field in combination with the longitudinal magnetic field $B_x$ generates and sustains an ${\bf{E}} \times {\bf{B}}$ electron current directed along the $z$-axis. The corresponding current density is shown in the central panel of Fig.~\ref{Figure_sheath}. The generated current effectively leads to a diamagnetic response of the plasma plume to the applied magnetic field. As shown in the right panel of Fig.~\ref{Figure_sheath}, the total longitudinal magnetic field inside the expanding plasma plume is appreciably reduced compared to $B_0$. On the other hand, the same current causes a noticeable enhancement of $B_x$ on the periphery of the plume. As a result, the energetic electrons in the plume remain laterally well confined. 

The electron confinement significantly changes the topology of the sheath electric field that accelerates protons, as evident from Fig.~\ref{Figure_field}. The left panel shows a conventional bell-shaped sheath that is typical for the TNSA regime without an applied magnetic field~\cite{Romagnani2005}. The right panel shows a line-out along the axis of the laser pulse that heated the electrons ($y = 0$ $\mu$m). The electric field has a peak at the edge of the expanding plasma plume, while its amplitude is significantly reduced inside the plasma. The middle panel shows the amplitude of the electric field and the corresponding vector field at the rear side of the target in the presence of a 1.5 kT magnetic field. There are two distinct features that clearly distinguish this regime of plasma expansion from that without the applied magnetic field: 1) electrons generate a focusing sheath electric field for the protons at the edge of the plasma plume and 2) there is a significant longitudinal electric field in the plasma plume.

Figure~\ref{Figure_ions} shows snapshots of proton densities produced by the sheath electric fields 1.7 ps after the laser-target interaction. At this point, a considerable amount of energy absorbed by the electrons during the laser-target interaction has been already transferred to the protons. Without the magnetic field, the protons have been accelerated off the entire rear surface of the target, whereas the proton beam is laterally limited in the presence of the magnetic field. This implies that the hot electron population does not spread laterally while expanding in the longitudinal direction. The reduced lateral heat outflow allows the electrons to sustain stronger proton acceleration, similar to what was reported by S. Buffechoux et al.~\cite{Buffechoux2010} and T. Toncian et al.~\cite{Toncian2011} for targets with reduced surface area. This aspect becomes already evident in the right panel of Fig.~\ref{Figure_field}, where the peak of the electric field that corresponds to the plasma edge and thus to the most energetic protons is further from the target for $B_0 = 1.5$ kT. The most energetic protons have higher energies at $B_0 = 1.5$ kT, since they were able to travel further. The right panel of Fig.~\ref{Figure_edf} confirms that not only the most energetic protons whose energies are 50\% higher, but the entire proton energy spectrum benefits from the reduced lateral heat outflow in the presence of the magnetic field.

\begin{figure*}[bth]
	\centering
	\includegraphics[width=1.9\columnwidth]{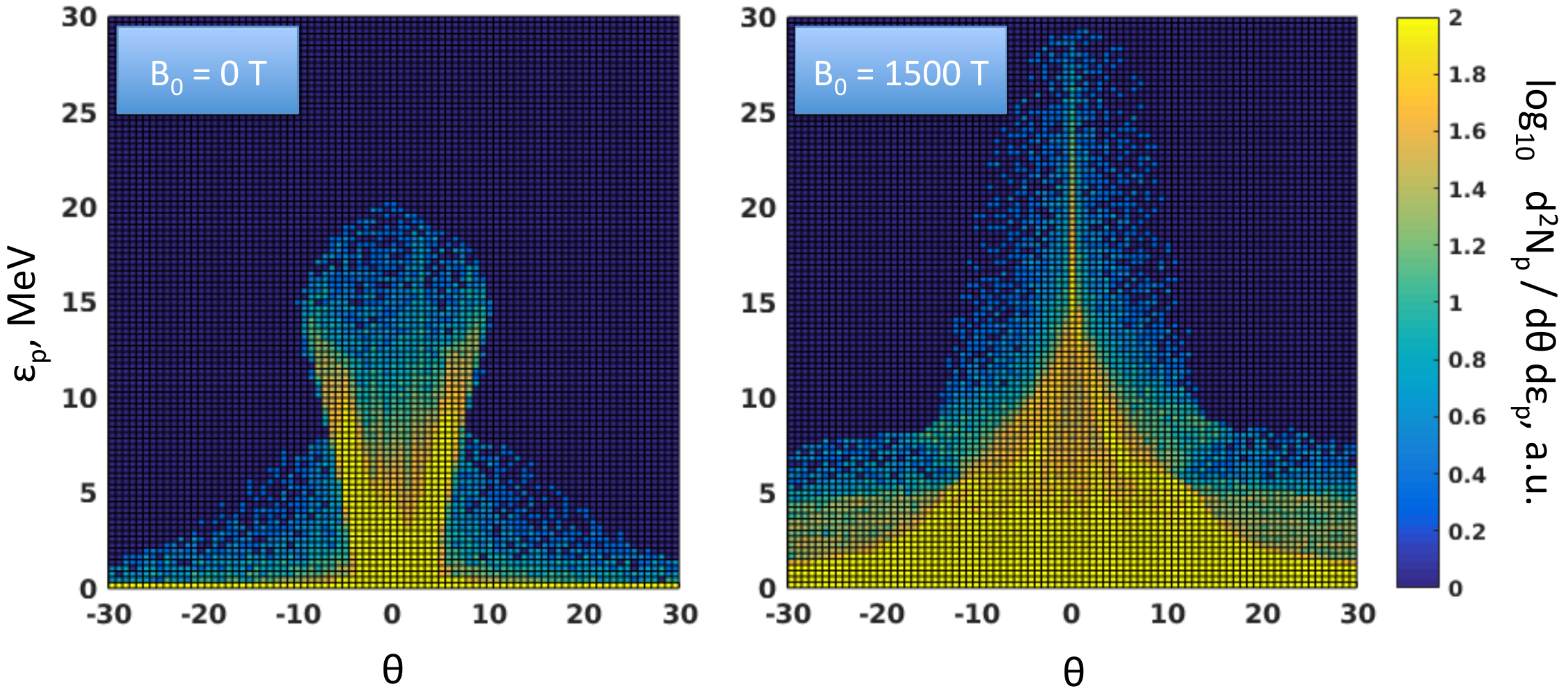} 
	  \caption{Proton distribution as a function of energy and angle at $t = t_0 + 1.7$ ps for the case with (right) and without (left) the applied magnetic field. The angle $\theta$ is the angle between the the proton momentum and the $x$-axis.} \label{Figure_p_theta}
\end{figure*}

A strong kT-level magnetic field not just enhances the energy spectrum of accelerated protons, but it also significantly reduces its divergence. Figure~\ref{Figure_p_theta} shows snapshots of the proton distribution as a function of proton energy $\varepsilon_i$ and the angle $\theta$ between the proton momentum and the $x$-axis. The snapshot are taken at 1.7 ps after the laser-target interaction, so a sum over all the angles for these distributions yields the proton energy distributions shown in the right panel of Fig.~\ref{Figure_edf} for the cases with and without the applied magnetic field. The focusing electric field structure shown in the middle panel of Fig.~\ref{Figure_field} generates a population of energetic protons ($\varepsilon_i > 15$ MeV) that is sharply peaked at $\theta = 0$ for $B_0 = 1.5$ kT, as compared to the case without the applied magnetic field. 


%
%


\section{Summary} \label{Sec-Summary}

We have performed 2D PIC simulations to examine the role of a static applied magnetic field on proton acceleration in laser-irradiated targets. We find that, for a laser peak amplitude of $a_0 = 4$, a kT-level  magnetic field can significantly suppress transverse transport of hot electron in the target. The suppression significantly enhances the resulting proton energy spectrum over a wide range of energies and increases the cutoff proton energy by approximately 50\%. The applied magnetic field also significantly reduces the angular spread of the proton beam by inducing a focusing sheath electric field for the protons. The total conversion rate of the incoming laser energy into the multi-MeV proton beam is 2.4 times higher than in the case without the applied magnetic field.

\section*{Acknowledgments}
The work by A. A. and T. T. was supported by U.S. Department of Energy - National Nuclear Security Administration Cooperative Agreement No. DE-NA0002008. Simulations for this work were performed using the EPOCH code (developed under UK EPSRC grants EP/G054940/1, EP/G055165/1 and EP/G056803/1) and HPC resources provided by the Texas Advanced Computing Center at The University of Texas. 





\section*{References}
\begin{thebibliography}{10}

\bibitem{Patel2003} 
P. K. Patel, A. J. Mackinnon, M. H. Key, T. E. Cowan, M. E. Foord, M. Allen, D. F. Price, H. Ruhl, P. T. Springer, and R. Stephens, ``{\it{Isochoric heating of solid-density matter with an ultrafast proton beam}}'', Phys. Rev. Lett., {\bf{91}}, 125004 (2003).

\bibitem{Borghesi2002} 
M. Borghesi, S. Bulanov, D. H. Campbell, R. J. Clarke, T. Z. Esirkepov, M. Galimberti, L. A. Gizzi, A. J. MacKinnon, N. M. Naumova, F. Pegoraro, H. Ruhl, A. Schiavi, and O. Willi, ``{\it{Macroscopic evidence of soliton formation in multiterawatt laser-plasma interaction}}'', Phys. Rev. Lett. {\bf{88}}, 135002 (2002).

\bibitem{Roth2001} 
M. Roth, T. E. Cowan, M. H. Key, S. P. Hatchett, C. Brown, W. Fountain, J. Johnson, D. M. Pennington, R. A. Snavely, S. C. Wilks, K. Yasuike, H. Ruhl, F. Pegoraro, S. V. Bulanov, E. M. Campbell, M. D. Perry, and H. Powell, ``{\it{Fast ignition by intense laser-accelerated proton beams}}'', Phys. Rev. Lett. {\bf{86}}, 436 (2001).
 
\bibitem{Bulanov2002}  
 S. V. Bulanov, T. Z. Esirkepov, F. F. Kamenets, Y. Kato, A. V. Kuznetsov, K. Nishihara, F. Pegoraro, T. Tajima, and V. S. Khoroshkov, ``{\it{Generation of high-quality charged particle beams during the acceleration of ions by high-power laser radiation}}'', Plasma Physics Reports {\bf{28}}, 975 (2002).
 
\bibitem{Spencer2001}
I. Spencer, K. W. D. Ledingham, R. P. Singhal, T. McCanny, P. McKenna, E. L. Clark, K. Krushelnick, M. Zepf, F. N. Beg, M. Tatarakis, A. E. Dangor, P. A. Norreys, R. J. Clarke, R. M. Allott, and I. N. Ross, ``{\it{Laser generation of proton beams for the production of short-lived positron emitting radioisotopes}}'', Nuclear Instruments and Methods In Physics Research Section B-Beam Interactions With Materials and Atoms {\bf{183}}, 449 (2001).

\bibitem{Gao2016} 
L. Gao, H. Ji, G. Fiksel, W. Fox, M. Evans, and N. Alfonso, ``{\it{Ultrafast proton radiography of the magnetic fields generated by a laser-driven coil current}}'', Phys. Plasmas. {\bf{23}}, 043106 (2016).


\bibitem{Law2016}
K. F. F. Law, M. Bailly-Grandvaux, A. Morace, S. Sakata, K. Matsuo, S. Kojima, S. Lee, X. Vaisseau, Y. Arikawa, A. Yogo, K. Kondo, Z. Zhang, C. Bellei, J. J. Santos, S. Fujioka, and H. Azechi, ``{\it{Direct measurement of kilo-tesla level magnetic field generated with laser-driven capacitor-coil target by proton deflectometry}}'', Appl. Phys. Lett. {\bf{108}}, 091104 (2016).

\bibitem{Wilks2001}
S. C. Wilks, A. B. Langdon, T. E. Cowan, M. Roth, M. Singh, S. Hatchett, M. H. Key, D. Pennington, A. MacKinnon, and R. A. Snavely, ``{\it{Energetic proton generation in ultra-intense laser-solid interactions}}'', Phys. Plasmas. {\bf{8}}, 542 (2001).


\bibitem{Peebles2016}
J. Peebles, C. McGuffey, C. M. Krauland, L. C. Jarrott, A. Sorokovikova, M. S. Wei, J. Park, H. Chen, H. S. McLean, C. Wagner, M. Spinks, E. W. Gaul, G. Dyer, B. M. Hegelich, M. Martinez, M. Donovan, T. Ditmire, S. I. Krasheninnikov, and F. N. Beg, ``{\it{Impact of pre-plasma on fast electron generation and transport from short pulse, high intensity lasers}}'', Nucl. Fusion {\bf{56}}, 016007 (2016).

\bibitem{Sorokovikova2016}
A. Sorokovikova, A. V. Arefiev, C. McGuffey, B. Qiao, A. P. L. Robinson, M. S. Wei, H. S. McLean, and F. N. Beg, ``{\it{Generation of superponderomotive electrons in multipicosecond interactions of kilojoule laser beams with solid-density plasmas}}'', Phys. Rev. Lett. {\bf{116}}, 155001 (2016).

\bibitem{Arefiev2016}
A. V. Arefiev, V. N. Khudik, A. P. L. Robinson, G. Shvets, L. Willingale, and M. Schollmeier, ``{\it{Beyond the ponderomotive limit: Direct laser acceleration of relativistic electrons in sub-critical plasmas}}'', Phys. Plasmas. {\bf{23}}, 056704 (2016).

\bibitem{Gurevich} 
A. V. Gurevich, L. V. Pariiska, and L. P. Pitaevskii, ``{\it{Self-similar motion of rarefied plasma}}'', Sov. Phys. JETP {\bf{22}}, 449 (1966).
 
 \bibitem{Mora2003} 
P. Mora, ``{\it{Plasma expansion into a vacuum}}'', Phys. Rev. Lett. {\bf{90}}, 185002 (2003).

\bibitem{Mora2005}  
P. Mora, ``{\it{Thin-foil expansion into a vacuum}}'', Phys. Rev. E, {\bf{72}}, 056401 (2005).

\bibitem{Romagnani2005} 
L. Romagnani, J. Fuchs, M. Borghesi, P. Antici, P. Audebert, F. Ceccherini, T. Cowan, T. Grismayer, S. Kar, A. Macchi, P. Mora, G. Pretzler, A. Schiavi, T. Toncian, and O. Willi, ``{\it{Dynamics of electric fields driving the laser acceleration of multi-MeV protons}}'', Phys. Rev. Lett. {\bf{95}}, 195001 (2005).

\bibitem{Psikal2008} 
J. Psikal, V. T. Tikhonchuk, J. Limpouch, A. A. Andreev, and A. V. Brantov, ``{\it{Ion acceleration by femtosecond laser pulses in small multispecies targets}}'', Phys. Plasmas {\bf{15}}, 053102 (2008).

\bibitem{Buffechoux2010}
S. Buffechoux, J. Psikal, M. Nakatsutsumi, L. Romagnani, A. Andreev, K. Zeil, M. Amin, P. Antici, T. Burris-Mog, A. Compant-La-Fontaine, E. d?Humieres, S. Fourmaux, S. Gaillard, F. Gobet, F. Hannachi, S. Kraft, A. Mancic, C. Plaisir, G. Sarri, M. Tarisien, T. Toncian, U. Schramm, M. Tampo, P. Audebert, O. Willi, T. E. Cowan, H. Pepin, V. Tikhonchuk, M. Borghesi, and J. Fuchs, ``{\it{Hot electrons transverse refluxing in ultraintense laser-solid interactions}}'', Phys. Rev. Lett. {\bf{105}}, 015005 (2010).

\bibitem{Toncian2011}
T. Toncian, M. Swantusch, M. Toncian, O. Willi, A. A. Andreev, and K. Y. Platonov, ``{\it{Optimal proton acceleration from lateral limited foil sections and different laser pulse durations at relativistic intensity}}'', Phys. Plasmas {\bf{18}}, 043105 (2011).

\bibitem{Margarone2012} 
D. Margarone, O. Klimo, I.J. Kim, J. Prokupek, J. Limpouch, T.M. Jeong, T. Mocek, J. Pkal, H.T. Kim, J. Proka, K.H. Nam, L. Tolcov, I.W. Choi, S.K. Lee, J.H. Sung, T.J. Yu, and G. Korn, ``{\it{Laser-driven proton acceleration enhancement by nanostructured foils}}'', Phys. Rev. Lett. {\bf{109}}, 234801 (2012).

\bibitem{Kar2011} 
S. Kar, K. Markey, M. Borghesi, D. C Carroll, P. McKenna, D. Neely, M. N. Quinn, and M. Zepf, ``{\it{Ballistic focusing of polyenergetic protons driven by petawatt laser pulses}}'', Phys. Rev. Lett. {\bf{106}}, 225003 (2011).
 
\bibitem{Offermann2011} 
D. T. Offermann, K. A. Flippo, J. Cobble, M. J. Schmitt, S. A. Gaillard, T. Bar- tal, D. V. Rose, D. R. Welch, M. Geissel, and M. Schollmeier, ``{\it{Characterization and focusing of light ion beams generated by ultra-intensely irradiated thin foils at the kilojoule scale}}'', Phys. Plasmas {\bf{18}}, 056713 (2011).




\bibitem{Daido}  
H. Daido, F. Miki, K. Mima, M. Fujita, K. Sawai, H. Fujita, Y. Kitagawa, S. Nakai, and C. Yamanaka, ``{\it{Generation of a strong magnetic field by an intense CO$_2$ laser pulse}}'', Phys. Rev. Lett. {\bf{56}}, 846 (1986).

\bibitem{Daido87}
H. Daido, K. Mima, F. Miki, M. Fujita, Y. Kitagawa, S. Nakai, and C. Yamanaka, ``{\it{Ultrahigh pulsed magnetic field produced by a CO$_2$ laser}}'', J. J. Applied Physics {\bf{26}}, 1290 (1987). 

\bibitem{Fujioka} 
S. Fujioka, Z. Zhang, K. Ishihara, K. Shigemori, Y. Hironaka, T. Johzaki, A. Sunahara, N. Yamamoto, H. Nakashima, T. Watanabe, H. Shiraga, H. Nishimura, and H. Azechi, ``{\it{Kilotesla magnetic field due to a capacitor-coil target driven by high power laser}}'', Scientific Rep. {\bf{3}}, 1170 (2013).


\bibitem{Mima2016}
K. Mima, T. Johzaki, J. Honrubia, H. Nagatomo, T. Taguchi, A. Sunahara, H. Sakagami, S. Fujioka and G. Logan, Magnetized Fast ignition (MFI) and Laser Plasma Interactions in Strong Magnetic Field, Journal of Physics: Conference Series 688 (2016) 012066.



\bibitem{Arber2015}
T. D. Arber, K. Bennett, C. S. Brady, A. Lawrence-Douglas, M. G. Ramsay, N. J. Sircombe, P. Gillies, R. G. Evans, H. Schmitz, A. R. Bell,  and C. P. Ridgers, ``{\it{Contemporary particle-in-cell approach to laser-plasma modelling}}'', Plasma Phys. Control. Fusion {\bf{57}}, 113001 (2015).

\bibitem{Schollmeier2015}
M. Schollmeier, A. B. Sefkow, M. Geissel, A. V. Arefiev, K. A. Flippo, S. A. Gaillard, R. P. Johnson, M. W. Kimmel, D. T. Offermann, P. K. Rambo, J. Schwarz, and T. Shimada, ``{\it{Laser-to-hot-electron conversion limitations in relativistic laser matter interactions due to multi-picosecond dynamics}}'', Phys. Plasmas {\bf{22}}, 043116 (2015).

\bibitem{Arefiev2015}
A. V. Arefiev, A. P. L. Robinson, and V. N. Khudik, ``{\it{Novel aspects of direct laser acceleration of relativistic electrons}}'', J. Plasma Phys. {\bf{81}}, 475810404 (2015).


\end{thebibliography}
\end{document}